\documentclass[runningheads]{cl2emult}

\usepackage{makeidx}  
\usepackage{graphicx} 
\usepackage{multicol} 
\makeindex            

\begin{document}
\title*{Non-linearity and Non-Gaussianity through Phase Information}
\toctitle{Non-linearity and Non-Gaussianity through Phase
Information}
%
%
\titlerunning{Phase Information}
%
\author{Peter Coles\inst{1}
\and Lung-Yih Chiang\inst{2}}
\authorrunning{Peter Coles and Lung-Yih Chiang}
%
%
\institute{School of Physics \& Astronomy, University of
Nottingham, University Park, Nottingham NG7 2RD, United Kingdom
 \and Theoretical Astrophysics Center,
     Juliane Maries Vej 30, DK-2100 Copenhagen, Denmark}

\maketitle              

\begin{abstract}
In the standard picture of structure formation, initially
random-phase fluctuations are amplified by non-linear
gravitational instability to produce a final distribution of mass
which is highly non-Gaussian and has highly coupled Fourier
phases. Second-order statistics, such as the power spectrum, are
blind to this kind of phase association. We discuss the
information contained in the phases of cosmological density
fluctuations and their possible use in statistical analysis tools.
In particular, we show how the bispectrum measures a particular
form of phase association called quadratic phase coupling, show
how to visualise phase association using colour models. These
techniques offer the prospect of more complete tests of initial
non-Gaussianity than those available at present.
\end{abstract}

\section{Introduction}

The local Universe displays a rich hierarchical pattern of galaxy
clusters and superclusters~\cite{lasc}. The early Universe,
however, was almost smooth, with only slight ripples seen in the
cosmic microwave background radiation~\cite{Cobe}. Models of the
evolution of structure link these observations through the effect
of gravity, because the small initially overdense fluctuations
attract additional mass as the Universe expands~\cite{p80}. During
the early stages, the ripples evolve independently, like linear
waves on the surface of deep water. As the structures grow in
mass, they interact with other in non-linear ways, more like waves
breaking in shallow water. Cosmic structure can be characterized
by phase correlations associated with these non-linear
interactions, but this information is missed by standard analysis
techniques such as the power spectrum. In order to do justice to
the large data sets about to become available, it is important to
design techniques sensitive to the fine details of cosmic
structure they will reveal. Here we report a method of quantifying
phase information \cite{cc} and suggest how this information may
be exploited to build novel statistical descriptors that can be
used to mine the sky more effectively than with standard methods.

\section{Fourier Description of Cosmological Density Fields}
In most popular versions of the ``gravitational instability''
model for the origin of cosmic structure, particularly those
involving cosmic inflation~\cite{gp}, the initial fluctuations
that seeded the structure formation process form a Gaussian random
field~\cite{bbks}. Deviations from uniformity, expressed in terms
of the density contrast $\delta({\bf x})$ defined by $ \delta
({\bf x}) =[\rho({\bf x})-\rho_0]/\rho_0$,
 where $\rho_0$ is the average density and
$\rho({\bf x})$ is the local matter density. Because the initial
perturbations evolve linearly, it is useful to expand $\delta({\bf
x})$  as a Fourier superposition of plane waves:
\begin{equation}
\delta ({\bf x}) =\sum \tilde{\delta}({\bf k}) \exp(i{\bf k}\cdot
{\bf x}).
\end{equation}
The Fourier transform $\tilde{\delta}({\bf k})$ is complex and
therefore possesses both amplitude $|\tilde{\delta} ({\bf k})|$
and phase $\phi_{\bf k}$ where
\begin{equation}
\tilde{\delta}({\bf k})=|\tilde{\delta} ({\bf
k})|\exp(i\phi_{\bf_k}). \label{eq:fourierex}
\end{equation}
Gaussian random fields possess Fourier modes whose real and
imaginary parts are independently distributed. In other words,
they have phase angles $\phi_k$ that are independently distributed
and uniformly random on the interval $[0,2\pi]$. When fluctuations
are small, i.e. during the linear regime, the Fourier modes evolve
independently and their phases remain random. In the later stages
of evolution, however, wave modes begin to couple
together~\cite{p80}. In this  regime the phases become non-random
and the density field becomes highly non--Gaussian. Phase coupling
is therefore a key consequence of nonlinear gravitational
processes if the initial conditions are Gaussian and a potentially
powerful signature to exploit in statistical tests of this class
of models.

A graphic demonstration of the importance of phases in patterns
generally is given in Fig 1.
\begin{figure}
\centering
\includegraphics[width=0.45\textwidth]{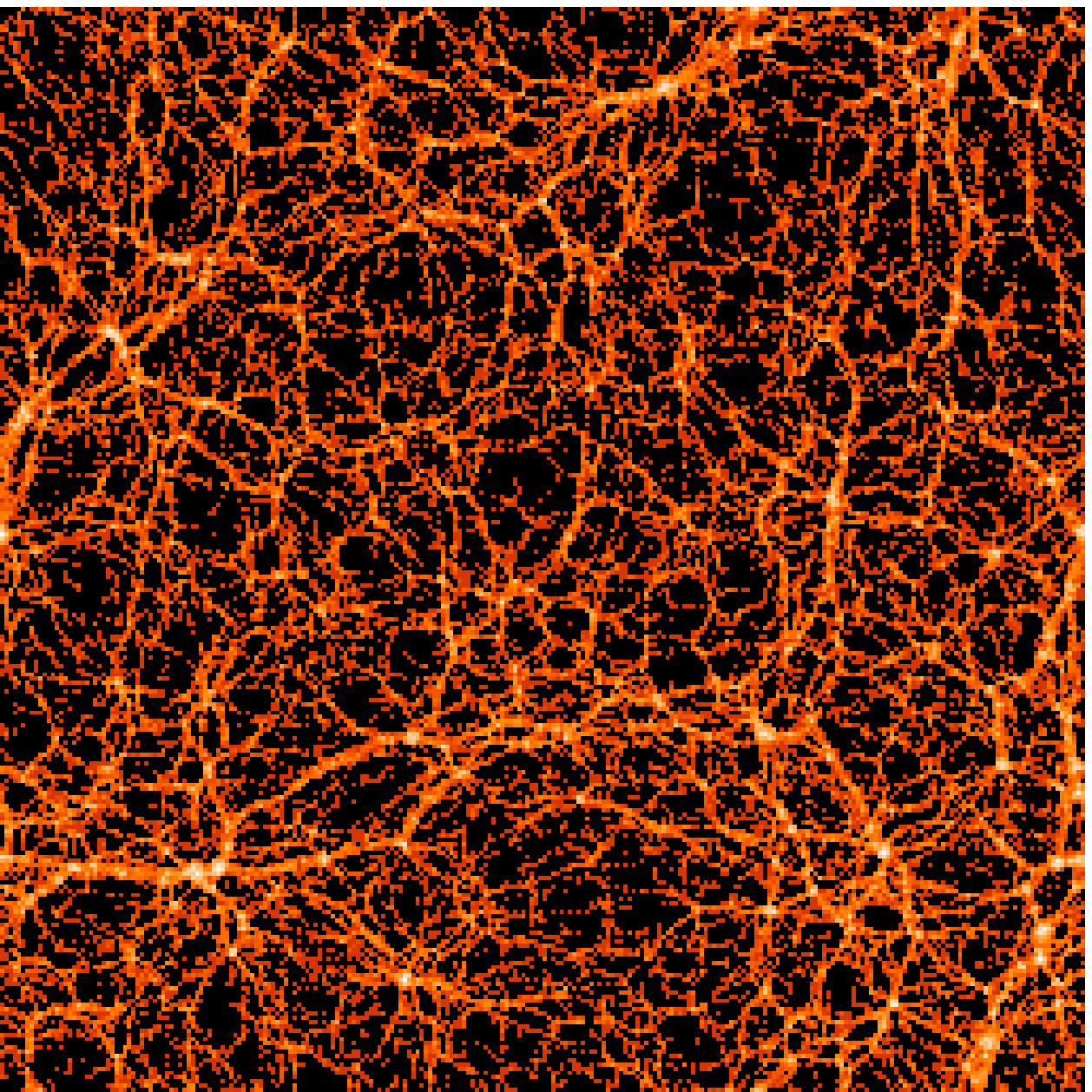}
\includegraphics[width=0.45\textwidth]{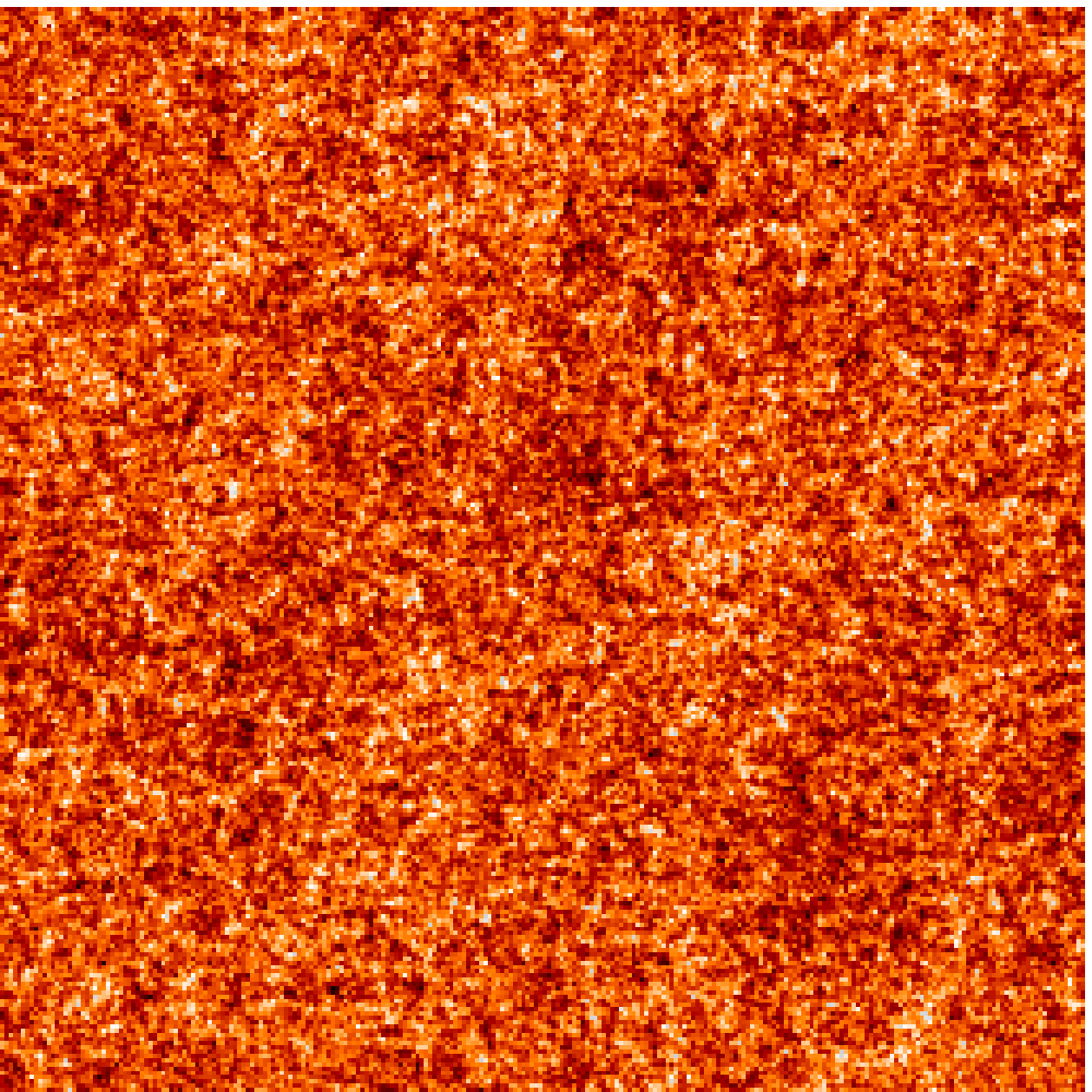}
\caption[]{Numerical simulation of galaxy clustering (left)
together with a version generated  randomly reshuffling the phases
between Fourier modes of the original picture (right).}
\label{eps1}
\end{figure}
Since the amplitude of each Fourier mode is unchanged in the phase
reshuffling operation, these two pictures have exactly the same
power-spectrum, $P(k)\propto|\tilde{\delta}({\bf k})|^2$. In fact,
they have more than that: they have exactly the same amplitudes
for all ${\bf k}$. They also have totally different morphology.
The evident shortcomings of $P(k)$ can be partly ameliorated by
defining higher-order quantities such as the
bispectrum~\cite{p80,mvh,scf,vwhk} or correlations of
$\tilde{\delta}({\bf k})^2$ \cite{stir}.

\section{The Bispectrum and Phase Coupling}
The bispectrum and higher-order polyspectra vanish for Gaussian
fields, but in a non-Gaussian field they may be non-zero. The
usefulness of these and related quantities therefore lies in the
fact that they encode some information about non-linearity and
non-Gaussianity.  To understand the relationship between the
bispectrum and Fourier phases, it is very helpful to consider the
following toy examples. Imagine a simple density field defined in
one spatial dimension that consists of the superposition of two
cosine components:
\begin{equation}
\delta(x) = A_1 \cos (\lambda_1x+\phi_1)+A_2 \cos (\lambda_2x +
\phi_2). \label{eq:toy1}
\end{equation}
The generalisation to several spatial dimensions is trivial. The
phases $\phi_1$ and $\phi_2$ are random and $A_1$ and $A_2$ are
constants. We can simplify the following by introducing a new
notation
\begin{equation} \delta(x)= A_1 \left(\begin{array}{c} \lambda_1
\\
\phi_1 \end{array} \right) + A_2 \left(\begin{array}{c} \lambda_2
\\ \phi_2 \end{array} \right). \label{eq:toy2} \end{equation}
Clearly this example displays no phase correlations. Now consider
a new field obtained from the example (\ref{eq:toy1}) through the
non-linear transformation
\begin{equation}
\delta(x)\mapsto \delta(x)+\epsilon \delta^2(x), \label{eq:nl}
\end{equation}
where $\epsilon$ is a constant parameter. Equation (\ref{eq:nl})
may be thought of as a very phenomenological representation of a
perturbation series, with $\epsilon$ controlling the level of
non-linearity. Using the same notation as equation
(\ref{eq:toy2}), the new field $\delta(x)$ can be written
\begin{eqnarray}
\delta(x) & = & B_1 \left(\begin{array}{c} \lambda_1 \\ \phi_1
\end{array} \right) + B_2 \left(\begin{array}{c} \lambda_2 \\
\phi_2 \end{array} \right) + B_3 \left(\begin{array}{c} 2\lambda_1
\\ 2\phi_1 \end{array} \right) + B_4 \left(\begin{array}{c}
2\lambda_2 \\ 2\phi_2 \end{array} \right) + \nonumber \\ & & + B_5
\left(\begin{array}{c} \lambda_1 + \lambda_2 \\ \phi_1 + \phi_2
\end{array} \right) + B_6 \left(\begin{array}{c}
\lambda_1-\lambda_2 \\ \phi_1-\phi_2 \end{array} \right),
\label{eq:quadro}
\end{eqnarray}
where the $B_i$ are constants obtained from the $A_i$. Notice in
equation (\ref{eq:quadro}) that the phases follow the same kind of
harmonic relationship as the wavenumbers. This form of phase
association is termed {\em quadratic} phase coupling. It is this
form of phase relationship that appears in the bispectrum. To see
this, consider another two toy examples. First, model A,
\begin{equation}
\delta_{\rm A}(x) = \left(\begin{array}{c} \lambda_1 \\ \phi_1
\end{array} \right) + \left(\begin{array}{c} \lambda_2 \\ \phi_2
\end{array} \right) + \left(\begin{array}{c} \lambda_3 \\ \phi_3
\end{array} \right), \label{eq:modelA}
\end{equation}
in which $\lambda_3=\lambda_1+\lambda_2$ but in which $\phi_1$,
$\phi_2$ and $\phi_3$ are random; and
\begin{equation}
\delta_{\rm B}(x) = \left(\begin{array}{c} \lambda_1 \\ \phi_1
\end{array} \right) + \left(\begin{array}{c} \lambda_2 \\ \phi_2
\end{array} \right) + \left(\begin{array}{c} \lambda_3=
\lambda_1+\lambda_2 \\ \phi_3= \phi_1 +\phi_2 \end{array} \right).
\label{eq:modelB}
\end{equation}
Model A exhibits no phase association; model B displays quadratic
phase coupling. It is straightforward to show that $\langle
\delta_A\rangle=\langle \delta_B\rangle=0$. The autocovariances
are equal:
\begin{equation}
\xi_A(r)=\langle\delta_A (x) \delta_A(x+r)\rangle =\xi_B(r) =
\frac{1}{2} [\cos (\lambda_1r) + \cos (\lambda_2r) + \cos
(\lambda_3r)],
\end{equation}
as are the power spectra, demonstrating that second-order
statistics are blind to phase association.

The (reduced) three-point autocovariance function is
\begin{equation}
\zeta(r_1, r_2)= \langle
\delta(x)\delta(x+r_1)\delta(x+r_2)\rangle.
\end{equation}
For model A we get
\begin{equation}
\zeta_A(r_1, r_2)=0,
\end{equation}
whereas for model B it is
\begin{eqnarray}
\zeta_B(r_1,r_2) & = &  \frac{1}{4} \left[ \cos(\lambda_2 r_1 +
\lambda_1 r_2) + \cos(\lambda_3 r_1 - \lambda_1 r_2) +
\cos(\lambda_1 r_1 + \lambda_2 r_2)   \right.  \nonumber\\ & & +
\left. \cos(\lambda_3 r_1 - \lambda_2 r_2)  + \cos(\lambda_1 r_1 -
\lambda_3 r_2) + \cos(\lambda_2 r_1 - \lambda_3 r_2) \right].
\end{eqnarray}

The bispectrum, $B(k_1,k_2)$, is defined as the two-dimensional
Fourier transform of $\zeta$, so $B_A(k_1,k_2)=0$ trivially,
whereas $B_B(k_1, k_2)$ consists of a single spike located
somewhere in the region of $(k_1, k_2)$ space defined by $k_2\geq
0$, $k_1\geq k_2$ and $k_1+k_2\leq \pi$. If $\lambda_1\geq
\lambda_2$ then the spike appears at $k_1=\lambda_1$,
$k_2=\lambda_2)$. Thus the bispectrum  measures the phase coupling
induced by quadratic nonlinearities. To reinstate the phase
information order-by-order requires an infinite hierarchy of
polyspectra.

An alternative way of looking at this issue is to note that the
information needed to fully specify a non-Gaussian field to
arbitrary order (or, in a wider context, the information needed to
define an image resides in the complete set of Fourier
phases~\cite{opp}. Unfortunately, relatively little is known about
the behaviour of Fourier phases in the nonlinear regime of
gravitational clustering~\cite{ryden,sms,soda,Jain,Jain2,cc2}, but
it is of great importance to understand phase correlations in
order to design efficient statistical tools for the analysis of
clustering data.

\section{Visualizing and Quantifying Phase Information}
A vital first step on the road to a useful quantitative
description of phase information is to represent it
visually\cite{cc2}. In colour image display devices, each pixel
represents the intensity and colour at that position in the
image~\cite{thorn,fvd}. The quantitative specification of colour
involves three coordinates describing the location of that pixel
in an abstract colour space, designed to reflect as accurately as
possible the eye's response to light of different wavelengths. In
many devices this colour space is defined in terms of the amount
of Red, Green or Blue required to construct the appropriate tone;
hence the RGB colour scheme. The scheme we are particularly
interested in is based on three different parameters: Hue,
Saturation and Brightness. Hue is the term used to distinguish
between different basic colours (blue, yellow, red and so on).
Saturation refers to the purity of the colour, defined by how much
white is mixed with it. A saturated red hue would be a very bright
red, whereas a less saturated red would be pink. Brightness
indicates the overall intensity of the pixel on a grey scale. The
HSB colour model is particularly useful because of the properties
of the `hue' parameter, which is defined as a circular variable.
If the Fourier transform of a density map has real part $R$ and
imaginary part $I$ then the phase for each wavenumber, given by
$\phi=\arctan(I/R)$, can be represented as a hue for that pixel
using the colour circle~\cite{cc2}.

The pattern of phase information revealed by this method related
to the gravitational dynamics of its origin. For example in our
analysis of phase coupling~\cite{cc} we introduced a quantity
$D_k$, defined by
\begin{equation}
D_k\equiv\phi_{k+1}-\phi_{k},
\end{equation}
which measures the difference in phase of modes with neighbouring
wavenumbers in one dimension. We refer to $D_k$ as the phase
gradient. To apply this idea to a two-dimensional simulation we
simply calculate gradients in the $x$ and $y$ directions
independently. Since the difference between two circular random
variables is itself a circular random variable, the distribution
of $D_k$ should initially be uniform. As the fluctuations evolve
waves begin to collapse, spawning higher-frequency modes in phase
with the original~\cite{sz}. These then interact with other waves
to produce a non-uniform distribution of $D_k$. For  examples, see
\begin{verbatim}
http://www.nottingham.ac.uk/~ppzpc/phases/index.html.
\end{verbatim}

It is necessary to develop quantitative measures of phase
information that can describe the structure displayed in the
colour representations. In the beginning the phases $\phi_k$ are
random and so are the $D_k$ obtained from them. This corresponds
to a state of minimal information, or in other words maximum
entropy. As information flows into the phases  the information
content must increase and the entropy decrease. One way to
quantify this is by defining an information entropy on the set of
phase gradients. One constructs a frequency distribution, $f(D)$
of the values of $D_k$ obtained from the whole map. The entropy is
then defined as
\begin{equation} S(D)=-\int f(D)\log [f(D)] dD,
\end{equation}
where the integral is taken over all values of $D$, i.e. from $0$
to $2\pi$. The use of $D$, rather than $\phi$ itself, to define
entropy is one way of accounting for the lack of translation
invariance of $\phi$, a problem that was missed in previous
attempts to quantify phase entropy~\cite{pm}. A uniform
distribution of $D$ is a state of maximum entropy (minimum
information), corresponding to Gaussian initial conditions (random
phases). This maximal value of $S_{\rm max}=\log(2\pi)$ is a
characteristic of Gaussian fields. As the system evolves it moves
into to states of greater information content (i.e. lower
entropy). The scaling of $S$ with clustering growth displays
interesting properties~\cite{cc}, establishing an important link
between the spatial pattern and the physics driving clustering
growth.

\section{Discussion}

In fairly recent history,  cosmological data sets were sparse and
incomplete, and the statistical methods deployed to analyse them
were crude. Second-order statistics, such as $P(k)$ and $\xi(r)$,
are blunt instruments that throw away the fine details of the
delicate pattern of cosmic structure. These details lie in the
distribution of Fourier phases to which second-order statistics
are blind. It would not do justice to massively improved data if
effort were directed only to better estimates of these quantities.
Moreover, as we have shown, phase information provides a unique
fingerprint of gravitational instability developed from Gaussian
initial conditions (which have maximal phase entropy). Methods
such as those we have described above can therefore be used to
test this standard paradigm for structure formation. They can also
furnish direct tests of the presence of initial non-Gaussianity
~\cite{fmg,pvf,bt}. As the raw material is increasing in both
quality and quantity, it is time to refine our statistical
technology so that the subtle and precious artifacts previously
ignored can be both detected and extracted.

\clearpage
\addcontentsline{toc}{section}{Index}
\flushbottom
\printindex

\end{document}